\documentclass[aps,pre,reprint,amsmath,amssymb,superscriptaddress,showpacs,floatfix]{revtex4-2}%
\usepackage{graphicx}
\usepackage{dcolumn}
\usepackage{bm}
\usepackage[colorlinks, allcolors=blue]{hyperref}
\usepackage[usenames, dvipsnames]{color}
\usepackage{float}

\usepackage{lipsum}
\usepackage{titlesec}
\titlespacing\section{0pt}{12pt plus 4pt minus 4pt}{1pt plus 20pt minus 2pt}
\usepackage{xcolor}
\usepackage{tabularx}
\usepackage{amsmath}
\usepackage{comment}
\usepackage{afterpage}
\usepackage{placeins}
\usepackage{booktabs}
\usepackage{multirow}
\usepackage{array}
\usepackage{setspace}
\graphicspath{{Figs/}}
\usepackage{siunitx}
\usepackage{hhline}
\usepackage{xfrac}
\usepackage{float,graphicx}
\usepackage{mathtools}
\usepackage{listings}
\usepackage{amssymb}
\usepackage[normalem]{ulem}
\usepackage{titlesec}
\usepackage{amsfonts}
\usepackage[version=4]{mhchem}

\usepackage{epstopdf}
\catcode`@11
\def\seceqaa{\@addtoreset{equation}{section}
principles\def\theequation{A\arabic{equation}}}
\def\seceqbb{\@addtoreset{equation}{section}
\def\theequation{B\arabic{equation}}}
\def\seceqcc{\@addtoreset{equation}{section}
\def\theequation{C\arabic{equation}}}
\def\seceqdd{\@addtoreset{equation}{section}
\def\theequation{D\arabic{equation}}}
\def\seceqee{\@addtoreset{equation}{section}
\def\theequation{E\arabic{equation}}}
\def\seceqff{\@addtoreset{equation}{section}
\def\theequation{F\arabic{equation}}}
\def\seceqgg{\@addtoreset{equation}{section}
\def\theequation{G\arabic{equation}}}
\def\seceqhh{\@addtoreset{equation}{section}
\def\theequation{H\arabic{equation}}}
\catcode`@11

\begin{document}


\title{Spin-reorientation as a switch for electronic topology in van der Waals ferromagnets} 

\author{Satyabrata Bera}
\altaffiliation{These authors contributed equally to this work}
\affiliation{School of Physical Sciences, Indian Association for the Cultivation of Science, Jadavpur, Kolkata 700032, India}

\author{Sudipta Chatterjee}
\altaffiliation{These authors contributed equally to this work}
\affiliation{Department of Condensed Matter and Materials Physics, S. N. Bose National Centre for Basic Sciences, JD Block, Sector III, Salt Lake, Kolkata 700106, India}

\author{Suman Kalyan Pradhan}
\affiliation{Condensed Matter Physics Division, Saha Institute of Nuclear Physics, A
CI of Homi Bhabha National Institute 1/AF, Bidhannagar, Kolkata 700064, India}

\author{Subhadip Pradhan}

\affiliation{School of Physical Sciences, National Institute of Science
Education and Research Bhubaneswar, An OCC
of Homi Bhabha National Institute, Khurda Road, Jatni,
Odisha 752050, India} 
\author{Arnab Bera}
\affiliation{School of Physical Sciences, Indian Association for the Cultivation of Science, Jadavpur, Kolkata 700032, India}

\author{Sk Kalimuddin}
\affiliation{School of Physical Sciences, Indian Association for the Cultivation of Science, Jadavpur, Kolkata 700032, India}

\author{Ashis K. Nandy}
\affiliation{School of Physical Sciences, National Institute of Science
Education and Research Bhubaneswar, An OCC
of Homi Bhabha National Institute, Khurda Road, Jatni,
Odisha 752050, India}
\author{Mintu Mondal}
\email{sspmm4@iacs.res.in}
\affiliation{School of Physical Sciences, Indian Association for the Cultivation of Science, Jadavpur, Kolkata 700032, India}

\date{\today}

\begin{abstract}

The interplay between spin reorientation and topological electronic structure in two-dimensional (2D) van der Waals (vdW) ferromagnets (FM) is a core issue for understanding how magnetic anisotropy couples to non-trivial topological charge transport. Although spin-reorientation transitions (SRTs) are ubiquitous in 2D metallic FMs, their role in reshaping electronic-topology–driven thermodynamic and transport properties—especially magnetotransport and thermoelectric responses—has remained largely unexplored. Here, we address this issue in Fe$_4$GeTe$_2$ (F4GT), a room-temperature quasi-2D vdW FM, through comprehensive temperature-dependent magnetization, specific heat, magnetotransport, and thermoelectric measurements. Magnetization and specific-heat measurements establish a reorientation of the magnetic easy axis near $T_{\mathrm{SRT}} \sim 100$~K, in addition to ferromagnetic ordering at $T_C \sim 270$~K. Across the SRT, the Seebeck coefficient and the anisotropic magnetoresistance (AMR) display pronounced anomalies, suggesting a Fermi-surface reconstruction. The magnetoresistance shows a characteristic two-step field dependence: a low-field enhancement near the SRT associated with carrier scattering from canted spins and evolving domains, followed by a higher-field negative response as spin fluctuations are progressively suppressed. The concurrent sign change of the ordinary Hall coefficient $R_0$ and the sharp anomaly in the anomalous Hall resistivity ($\rho^{A}_{yx}$) across the SRT further suggests a temperature-driven modification of the underlying band topology. Furthermore, analysis of the anomalous Hall conductivity $\sigma^{A}_{yx}$ and scaling behavior of the $\rho^{A}_{yx}$ reveals that the Berry-curvature-driven anomalous Hall response below $T_{\mathrm{SRT}}$ is substantially modified above the transition. This study identifies spin reorientation as an efficient internal control parameter between distinct topological transport regimes in a 2D vdW FM, offering a symmetry-controlled route for engineering spin-polarized electronic states and domain-texture-driven functionalities. 
\end{abstract}

\maketitle

\section{INTRODUCTION}
Quasi-two-dimensional (2D) van der Waals (vdW) ferromagnets (FMs)---an versatile platform for studying the interaction between charge and spin degrees of freedom—have been at the forefront of condensed matter research owing to their unique physical properties and promise for spintronics applications ~\cite{Gong2017,Huang2017,Song2018,Klein2018,Zhong2017,RevModPhys.76.323,Wolf2001,sciadv.aay8912,PhysRevB.107.224422,PhysRevB.104.094405,Bader2010,PhysRevMaterials.2.051004,Sun2017}. In particular, 2D vdW FMs with a high Curie temperature ($T_C$) have been the epicenter in condensed matter physics over the past few years. In the majority of these FMs, the moments are typically oriented in a unique crystallographic direction as a consequence of spin-orbit coupling (SOC) \cite{lado2017origin}. However, in some cases, the spin direction reorients as a result of specific changes in physical parameters, such as temperature or pressure, leading to a spin-reorientation transition (SRT) in which the magnetic easy axis changes direction \cite{Rana2022,PhysRevB.93.134407,BERA2023170257}. While such transitions are routinely observed in 2D vdW FMs and identified through temperature-dependent magnetization and resistivity measurements, their impact on magnetotransport—particularly the associated underlying scattering mechanisms and their connection to the electronic topology—remains comparatively unexplored.

Within the family of 2D vdW FMs, the metallic compounds Fe$_n$GeTe$_2$ (with $n = 3, 4, 5$) have attracted significant attention owing to their relatively high Curie temperatures and rich physical properties~\cite{Bera2023,PhysRevB.107.224422,PhysRevB.108.115122,Adhikari2024,Bera2025}. Their magnetic and electronic properties are highly sensitive to chemical composition, particularly the Fe concentration, which governs both the magnetic transition temperature and the nature of the exchange interactions~\cite{sciadv.aay8912,May2019,Rana2022}. For example, the Curie temperature increases from approximately 220 K in Fe$_3$GeTe$_2$ (F3GT) to nearly 310 K in Fe$_5$GeTe$_2$ (F5GT)~\cite{Bera2023,May2019}. Beyond variations in magnetic ordering temperatures, the Fe$_n$GeTe$_2$ family exhibits pronounced differences in magnetocrystalline anisotropy and topological transport responses. The anomalous Hall effect (AHE) in bulk F3GT and Fe$_4$GeTe$_2$ (F4GT) originates predominantly from intrinsic Berry-phase contributions, whereas in F5GT both intrinsic and extrinsic mechanisms play an essential role~\cite{Kim2018,PhysRevB.108.115122,PhysRevMaterials.3.104401}. Furthermore, magnetic skyrmions have been experimentally observed in F3GT and F5GT, but to date, no such signatures have been reported in F4GT \cite{ding2019observation,schmitt2022skyrmionic}. Recently, topological flat bands have been reported in F5GT \cite{wu2025dichotomy}, whereas in F4GT, similar states have been theoretically predicted but remain experimentally unverified \cite{wang2023flat}. Another notable feature is the presence of SRTs, observed only in F4GT and F5GT among the Fe$_n$GeTe$_2$ members~\cite{sciadv.aay8912,Pal2024,May2019}. These findings collectively underscore pronounced differences in magnetocrystalline anisotropy energy and exchange interactions across this family of compounds.

In this work, we focus on F4GT, which has recently been reported to exhibit the highest anomalous Hall conductivity (AHC) and anomalous Hall factor within the Fe$_n$GeTe$_2$ family, attributed to Berry curvature arising from SOC–induced gapped nodal lines~\cite{ PhysRevB.108.115122}. F4GT also displays unusually high spin polarization and a pronounced, temperature-dependent magnetic anisotropy, characterized by a transition from in-plane to out-of-plane orientation upon cooling~\cite{sciadv.aay8912,Rana2022}. Density functional theory further predicts F4GT to be a promising half-metal with a tunneling magnetoresistance (MR) ratio approaching 500$\%$~\cite{halder2024half}. Moreover, a recent dynamical mean-field study also revealed that the peculiar magnetic anisotropy and multiple spin transitions in F4GT originate from strong local electronic correlations and site-dependent Fe-moment differentiation \cite{hasan2025dynamical}. Despite these notable characteristics, prior studies have focused primarily on the magnetic behavior near $T_C$, while the SRT remains comparatively unexplored, particularly from a magnetotransport perspective.

Here, we present a comprehensive investigation of the magnetic, thermodynamic, and magnetotransport properties of single-crystalline Fe$_4$GeTe$_2$ across its SRT. Magnetization and specific-heat measurements establish a well-defined reorientation of the magnetic easy axis near $T_{\mathrm{SRT}} \sim 100$~K, in addition to ferromagnetic ordering at $T_C \sim 270$~K. Complementary thermoelectric measurements reveal pronounced anomalies near $T_{\mathrm{SRT}}$, indicating a reconstruction of the Fermi surface. Detailed MR and Hall measurements uncover strong anisotropy and a crossover in the dominant carrier type across the transition. Furthermore, analysis of the AHC and the scaling behavior of the anomalous Hall resistivity $\rho^{A}_{yx}$ demonstrates that the intrinsic Berry-curvature-driven mechanism below $T_{\mathrm{SRT}}$ is strongly modified above the transition. These results highlight a close interplay between spin reorientation, Fermi-surface topology, and scattering mechanisms in this quasi-2D vdW FM, establishing spin reorientation as an effective internal control knob for topological electronic transport.

\begin{figure*}
\centering
\includegraphics[width=0.95\linewidth]{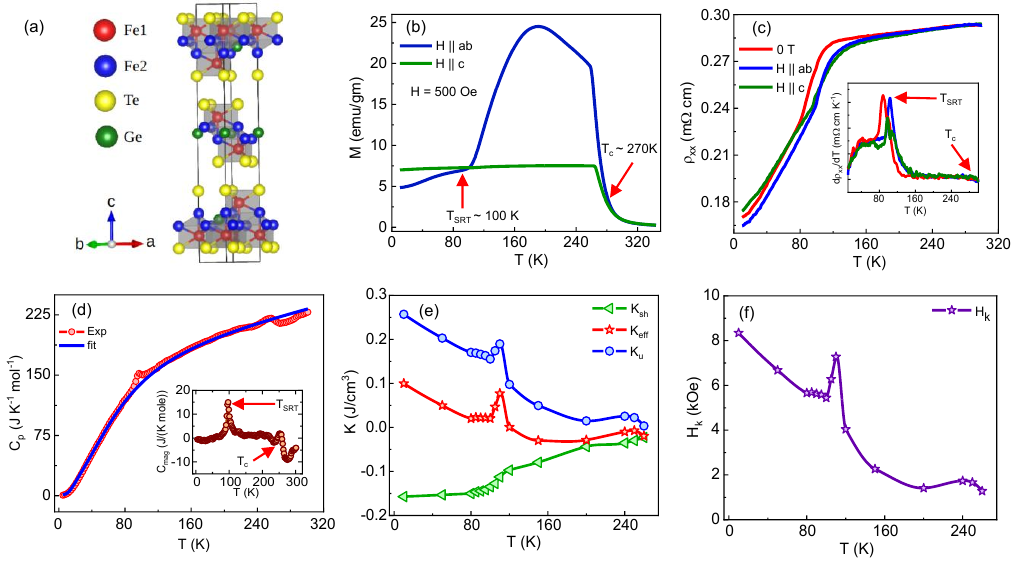}
\caption{\textbf{Crystal structure and temperature-dependent thermodynamic properties.} (a) Crystal structure of F4GT. (b) Temperature-dependent dc magnetization (M) in the zero-field-cooled mode for applied magnetic field H = 500 Oe in both directions. (c) The temperature dependence of longitudinal resistivity ($\rho_{xx}$) of F4GT single crystal at 0 T and 5 T with applied magnetic field parallel and perpendicular to $ab$-plane (current directions, I $\parallel$ $ab$). The inset shows the first-order derivative of $\rho_{xx}$ as a function of temperature. Arrows indicate the FM ordering temperature and the spin reorientation transition (SRT) temperature. (d) The temperature dependence of specific heat C$_p$. The inset shows the magnetic contribution to the heat capacity. (e) Measured magnetic anisotropy K$_{eff}$, calculated shape anisotropy K$_{sh}$, and resulting estimated uniaxial magnetocrystalline anisotropy K$_{u}$ as a function of temperature. (f) Calculated anisotropy field as a function of temperature.}
\label{fig1}
\vspace{-0.45cm}
\end{figure*}

\section{EXPERIMENTAL METHODS}
High-quality single crystals of F4GT were grown using the chemical vapor transport method with iodine (I$_2$) as the transport agent. Details of crystal growth, phase purity, and structural characterization are reported in our earlier publications~\cite{BERA2023170257,PhysRevB.108.115122}. Magnetization measurements were carried out using a Magnetic Property Measurement System (MPMS, Quantum Design). Thermoelectric (Seebeck coefficient) measurements were performed using a controlled temperature gradient technique implemented in a custom-built setup~\cite{Kalimuddin2026Hidden,Sruthi2022PRB}. A constant longitudinal temperature difference $\Delta T$ was established across the sample, and the resulting thermoelectric voltage $\Delta V$ was measured using a nanovoltmeter (Keithley 2182A). The Seebeck coefficient, S, was determined as $S = -\Delta V/\Delta T$~\cite{Kalimuddin2026Hidden}. Specific heat and magnetotransport measurements were conducted using a 9 T Physical Property Measurement System (PPMS, Quantum Design). Specific heat data were collected at zero applied magnetic field, and lattice contributions were analyzed using combined Debye–Einstein fits to extract the magnetic component. Both longitudinal ($\rho_{xx}$) and Hall ($\rho_{yx}$) resistivities were measured in a standard four-probe geometry with gold wires (25~$\mu$m diameter, Alfa Aesar) attached by silver paint (DuPont, 4929N). To eliminate artifacts arising from voltage probe misalignment, the MR and Hall resistivity data were symmetrized and antisymmetrized with respect to the magnetic field, respectively.

\section{RESULTS}

\subsection{Crystal structure and thermodynamic signatures of spin reorientation}

F4GT crystallizes in a centrosymmetric rhombohedral structure belonging to the $R\bar{3}m$ space group. The structure comprises Fe$_4$Ge slabs that are periodically sandwiched between vdW-bonded Te layers, forming a layered stacking along the c-axis, as shown in Fig. \ref{fig1}(a). Figure \ref{fig1}(b) shows the dc magnetization of the F4GT single crystal under a 500 Oe applied magnetic field along the \textit{ab} and \textit{c} axes. A sharp rise in \textit{M} on cooling marks the paramagnetic-to-ferromagnetic transition at $T_c$ $\sim$ 270 K, consistent with earlier reports \cite{Pal2024,PhysRevB.108.115122}. Between 270 and 165 K, $Fe_2$ (and some $Fe_1$) moments align with the field, giving a net FM state. On further cooling below 165 K, the increasing ordering of the remaining Fe$1$ moments leads to a gradual reduction of the net magnetization, and a second transition appears at $T_{SRT}$ $\sim$ 100 K, where the magnetization along \textit{c} slightly exceeds that in the \textit{ab} plane. This behavior signals a spin reorientation transition, associated with a change of the magnetic easy axis from the $ab$ plane to the $c$ axis upon cooling. We also measured the field-dependent magnetization at 10~K with the field applied both in the $ab$ plane ($H\parallel ab$) and along the $c$ axis ($H\parallel c$). In both geometries, $M(H)$ approaches saturation with a saturation magnetization $M_s \approx 1.95 ~\mu_B/\mathrm{Fe}$ (see Supplemental Material \cite{supply}). Notably, as shown in Fig.~S1(a), $M(H)$ for $H\parallel c$ reaches saturation at lower fields than for $H\parallel ab$, indicating that the magnetic easy axis is aligned along the $c$ axis at low temperature below $T_{SRT}$. 

The temperature dependence of the zero-field resistivity $\rho(T)$ for F4GT is shown in Fig.~\ref{fig1}(c). Two clear anomalies in the derivative $d\rho/dT$ (inset of Fig.~\ref{fig1}(c)) appear at $T_{\mathrm{SRT}} \approx 95~\mathrm{K}$ and $T_{c} \approx 270~\mathrm{K}$, consistent with previous reports~\cite{PhysRevB.108.115122}. To examine the influence of the magnetic field on charge transport, the resistivity measured at 5 T for fields applied parallel and perpendicular to the $ab$ plane (with current $I \parallel ab$) is plotted together with the zero-field data in Fig.~\ref{fig1}(c). A pronounced anomaly in $\rho(T,H)$ is observed in the vicinity of $T_{\mathrm{SRT}}$, whereas only a minimal change occurs near $T_{c}$. This strong field sensitivity near the SRT arises from enhanced scattering of itinerant carriers by canted spins and by reconfiguring magnetic domains, and it is significantly weaker across the ferromagnetic-paramagnetic transition, which we discuss further in the latter part of the paper. The temperature derivatives further reveal that the SRT anomaly shifts from $\sim 95$~K at zero field to 102~K for $H \parallel c$ and 105~K for $H \parallel ab$ under 5 T. The resistivity difference, $\rho_{\mathrm{diff}}(T) = \rho(5~\mathrm{T}) - \rho(0~\mathrm{T})$, exhibits its largest magnitude near $T_{\mathrm{SRT}}$ for both field orientations (see Fig. S2 of the Supplemental Material ~\cite{supply}). This highlights the emergence of a strong magnetotransport response at the reorientation temperature, where spin canting, thermally fluctuating domains, and field-induced spin reconfiguration maximize carrier scattering, leading to pronounced MR. Below $T \sim 60$~K, $\rho_{\mathrm{diff}}$ becomes positive for $H \parallel c$ while, remaining negative for $H \parallel ab$. This contrasting sign behavior reflects a substantial reduction of spin fluctuations and a rearrangement of magnetic domains at low temperature, both of which strongly couple to the conduction electrons and influence the MR~\cite{PhysRevB.85.094504,PhysRevB.100.214420}.

The temperature-dependent zero-field specific heat, $C_{p}(T)$, of the F4GT single crystal is shown in Fig.~\ref{fig1}(d). The total heat capacity can be written as 
$C_{p} = C_{\mathrm{elec}} + C_{\mathrm{lat}} + C_{\mathrm{mag}}$, where $C_{\mathrm{elec}} = \gamma T$ denotes the electronic contribution~\cite{PhysRevB.96.224426}. In the absence of an isostructural nonmagnetic analogue, $C_{\mathrm{mag}}$ was obtained by fitting the data with a combined Debye-Einstein model, since a pure Debye form does not adequately capture the experimental behavior. The fitting function is taken as~\cite{Martin1991}
\[
C_{p}^{\mathrm{fit}} = \gamma T + \alpha\,\mathcal{D}(T,\theta_{D}) + (1-\alpha)\,\mathcal{E}(T,\theta_{E}),
\]
where $\mathcal{D}(T,\theta_{D})$ and $\mathcal{E}(T,\theta_{E})$ are the Debye and Einstein functions, $\gamma$ is the Sommerfeld coefficient, $\theta_{D}$ and $\theta_{E}$ are the Debye and Einstein temperatures, and $\alpha$ is the corresponding weighting factor. The best fit yields $\gamma \approx 310~\mathrm{mJ~mol^{-1} K^{-2}}$, $\theta_{D} \approx 323~\mathrm{K}$, and $\theta_{E} \approx 87~\mathrm{K}$. The large value of $\gamma$ in F4GT indicates a strongly enhanced electronic density of states at the Fermi level, consistent with substantial effective-mass renormalization and pronounced electronic correlations~\cite{Deepali2024}. The magnetic contribution, obtained from $C_{\mathrm{mag}} = C_{p} - C_{p}^{\mathrm{fit}}$, is shown in the inset of Fig.~\ref{fig1}(d). A pronounced anomaly associated with the SRT is observed near $T_{\mathrm{SRT}}$ $\sim$ 100 K, while a broad maximum around 270 K corresponds to the ferromagnetic–paramagnetic transition.

\begin{figure*}
\centering
\includegraphics[width=1\textwidth]{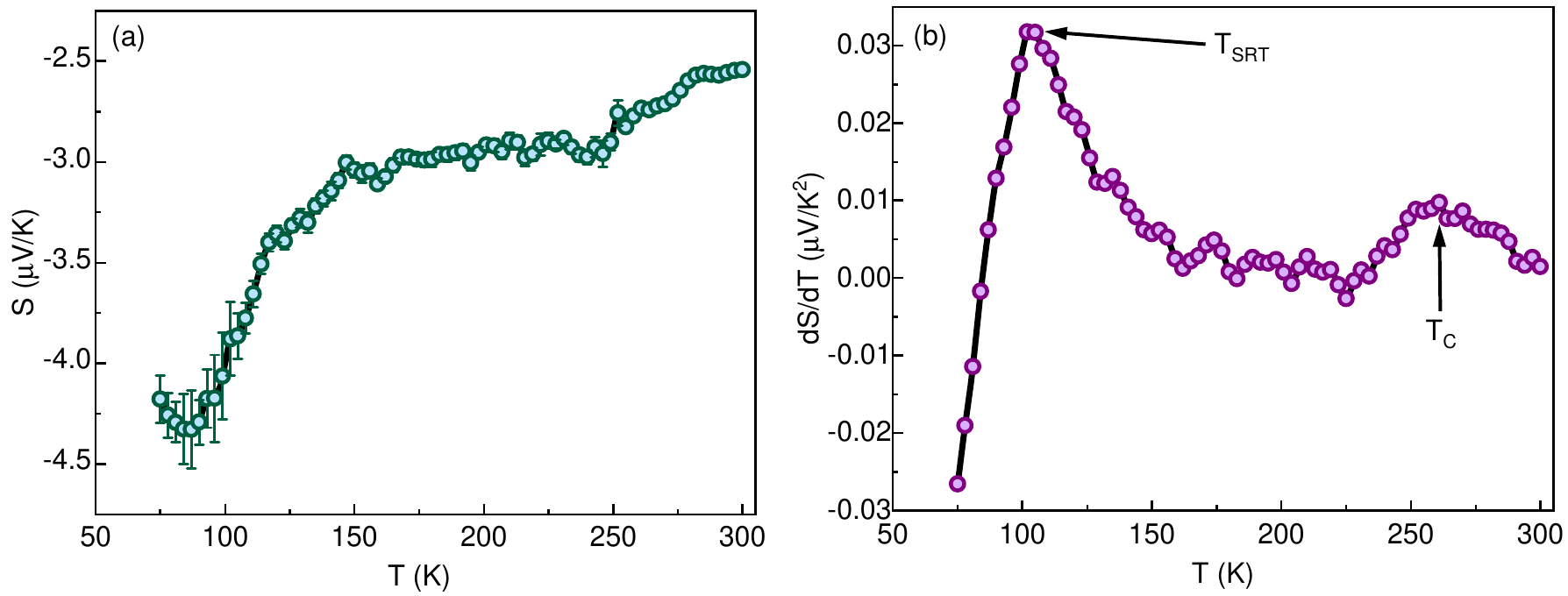}
\caption{\textbf{Thermoelectric response of Fe$_4$GeTe$_2$.}
(a) Temperature dependence of the Seebeck coefficient, $S(T)$, of Fe$_4$GeTe$_2$ single crystal. (b) Temperature derivative $dS/dT$, highlighting pronounced anomalies near the spin-reorientation transition and the ferromagnetic ordering temperature, indicative of a modification of the Fermi surface that manifested in transport response.}
\label{fig2}
\end{figure*}

The isothermal magnetization curves as a function of applied magnetic field at different temperatures, for fields applied along the in-plane (\(H \parallel ab\)) and out-of-plane (\(H \parallel c\)) directions, are shown in Fig. S1 (b-c) in the Supplemental Material \cite{supply}. The magnetizing energy for a given field along \(H \parallel ab\) and \(H \parallel c\) directions is obtained from the isothermal \(M\text{--}H\) data using the following equation \cite{PhysRevB.108.214417}:
\begin{equation}
 E_{H,i} = \int_{0}^{M_s} H\, dM.
\end{equation}
The effective magnetic anisotropy energy is then estimated as
\begin{equation}
 K_{\mathrm{eff}} = E_{H,\parallel} - E_{H,\perp},
\end{equation}
where \(E_{H,\parallel}\) and \(E_{H,\perp}\) are the out-of-plane and in-plane magnetizing energies, respectively. To determine the magnetocrystalline uniaxial anisotropy (\(K_u\)) of the system, the shape anisotropy $K_{sh} = -\frac{1}{2}\mu_0 N_d M_s^2$ is subtracted from the measured \(K_{\mathrm{eff}}\), i.e.,
\begin{equation}
 K_u = K_{\mathrm{eff}} - K_{\mathrm{sh}},
\end{equation}
where \(N_d\) and \(M_s\) are the demagnetizing factor and saturation magnetization of the sample, respectively. For a cuboid sample with in-plane dimensions \(u = v \simeq 1.2~\mathrm{mm}\) and thickness \(w \simeq 0.1~\mathrm{mm}\), the anisotropy difference between the in-plane and out-of-plane directions can be approximated as
\begin{equation}
 N_d = N_z - N_x = \frac{u - w}{u + 2w} \approx 0.785,
\end{equation}
using \(N_x = \frac{n}{2n+1}\) and \(N_z = \frac{1}{2n+1}\) with \(n = w/u\)~\cite{Sato1989}. The temperature dependence of the anisotropy energies \(K_u\), \(K_{\mathrm{eff}}\), and \(K_{\mathrm{sh}}\) is presented in Fig.~\ref{fig1}(e). The corresponding temperature-dependent anisotropy field,
\begin{equation}
 H_k = \frac{2K_u}{M_s},
\end{equation}
is shown in Fig.~\ref{fig1}(f). The uniaxial anisotropy exhibits a pronounced anomaly in the vicinity of the SRT. The magnetic anisotropy of F4GT is significantly smaller than that of other members of the 2D materials family; for example, Fe$_3$GeTe$_2$ exhibits an anisotropy energy density of $1.03~\mathrm{J/cm^3}$ \cite{Tan2018}. Such weak anisotropy is reflected in a comparatively small anisotropy field. Moreover, the near-zero-field linear dependence of $M(H)$ indicates that the bulk magnetization of F4GT is strongly influenced by fluctuations in the spin orientation and by the continuous evolution of magnetic domains under finite field and temperature \cite{Guo2021}.

\begin{figure*}
\centering
\includegraphics[width=1\textwidth]{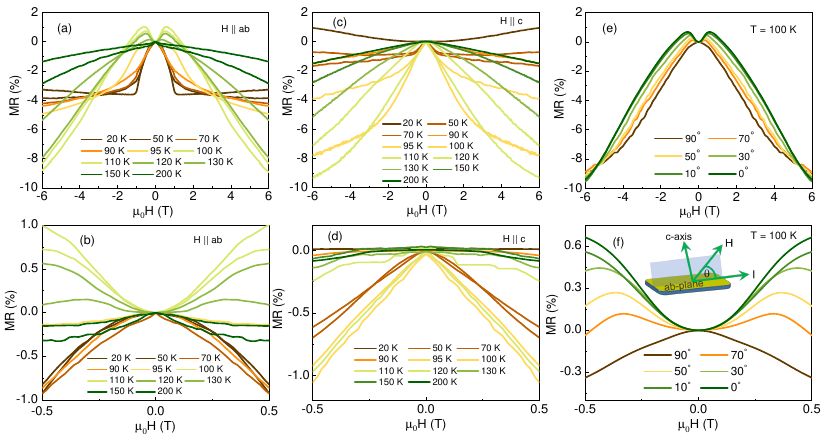}
\caption{\textbf{Magnetoresistance (MR) measurements.} MR of Fe$_4$GeTe$_2$ single crystal for the applied magnetic field (a) parallel to $ab$-plane and (c) along the $c$-axis, ranging from -6 to +6 T at a few representative temperatures in the range of 20-200 K. (b,d) Enlarged views of (a,c) around the low field. The magnetoresistance value is positive along H$\parallel$ab around $SRT$, but for the H$\parallel$c, the negative $MR$ value is higher than at other temperatures. (e) Resistivity as a function of H at various angles $\theta$ at 100 K. Angle-dependent MR (e) and enlarged views (f) around the low field. The MR value changes from positive to negative with increasing angle (a schematic diagram of sample orientation is shown in the inset).}
\label{fig3}
\end{figure*}

\subsection{Thermoelectric power and evidence of Fermi-surface reconstruction}

To further elucidate the electronic response associated with the SRT, we performed temperature-dependent Seebeck coefficient $S(T)$ measurements in F4GT. For a conventional metal in the weak-correlation limit, the thermopower can be described within semiclassical Boltzmann transport by the Mott relation \cite{cutler1969observation},
\begin{equation}
S(T)=-\frac{\pi^{2}}{3}\frac{k_{B}^{2}T}{e}
\left.\frac{\partial}{\partial \epsilon}\ln\!\big[\tau(\epsilon)\,v(\epsilon)\,S_{F}(\epsilon)\big]\right|_{\epsilon=\epsilon_{F}} .
\label{eq:mott}
\end{equation}
Here, $\tau$ is the transport relaxation time, $v$ is a characteristic Fermi velocity, and $S_{F}$ denotes the Fermi-surface area. When $\tau(\epsilon)$, $v(\epsilon)$, and $S_{F}(\epsilon)$ depend only weakly on energy near $\epsilon_F$, Eq.~(\ref{eq:mott}) yields the commonly observed approximately linear-in-$T$ behavior of $S(T)$. Conversely, appreciable energy dependence of any of these quantities can generate deviations from linearity. In particular, changes in $S_{F}$ associated with a Fermi-surface reconstruction can produce pronounced anomalies in the Seebeck coefficient \cite{chang2010nernst,Kalimuddin2026Hidden}.

The thermoelectric voltage $\Delta V$ generated by an applied longitudinal temperature gradient $\Delta T$ was recorded for different temperatures (see Section S3 of
the Supplemental Material ~\cite{supply}) and converted to $S(T)=-\Delta V/\Delta T$ as shown in Fig. \ref{fig2}(a). The resulting $S(T)$ remains negative throughout the measured temperature window ($\sim$75--300 K), with a small magnitude of only a few $\mu$V/K, consistent with metallic transport dominated by electron-like carriers. Importantly, $S(T)$ exhibits a distinct change in slope upon approaching $T_{\mathrm{SRT}}$, and the derivative $dS/dT$ displays a pronounced peak at $T_{\mathrm{SRT}}\sim 100$~K, together with an additional anomaly near the ferromagnetic ordering temperature $T_c\sim 270$~K [Fig. \ref{fig2}(b)]. These thermopower anomalies clearly indicate a modification of the low-energy transport landscape across the $T_{\mathrm{SRT}}$, consistent with changes in carrier scattering and/or a redistribution of contributions from different Fermi-surface sheets as the magnetic easy axis reorients.

\subsection{Magnetoresistance (MR)}

We now analyze the in-plane and out-of-plane MR at fixed temperatures. Figure~\ref{fig3}(a) shows the in-plane MR $(H \parallel ab)$ with the magnetic field applied parallel to the current ($H \parallel I$) over the range $-6~\mathrm{T}$ to $+6~\mathrm{T}$. The MR exhibits distinct temperature-dependent behavior. In the low-temperature regime ($20$--$70~\mathrm{K}$), MR decreases rapidly, reaching $\sim -3\%$ at $0.8~\mathrm{T}$, and then saturates. This evolution of ${\rm MR} $ is likely associated with a field-driven reconfiguration of the Fermi surface accompanying moment reorientation in this itinerant system. In contrast, in the intermediate temperature window ($95$--$130~\mathrm{K}$), MR shows a small positive contribution at low fields, reaching a maximum of $\sim 1\%$ near $0.5~\mathrm{T}$ before decreasing at higher fields [Fig. ~\ref{fig3}(b)]. This low-field enhancement is observed only in the vicinity of the SRT, i.e., between $100$ and $130~\mathrm{K}$, as highlighted by the magnified low-field view in Fig. ~\ref{fig3}(b). We attribute this behavior to the interaction between charge carriers and two distinct magnetic configurations of Fe moments and domains in this system. At all other temperatures, MR remains negative, consistent with the conventional response of an FM metal \cite{PhysRevB.60.3002,Tsunoda_2009,sudiptaCCGPhysRevB.107.125138}.

Figure ~\ref{fig3}(c) displays the out-of-plane MR $(H \parallel c)$ measured over the same field range, with the magnetic field applied perpendicular to the current ($H \perp I$). At low temperature and high field, the MR becomes positive, which we attribute to the suppression of fluctuating moments at low $T$ together with the conventional increase in resistivity from the Lorentz contribution \cite{Ke2020}. As the temperature increases, the magnitude of the negative MR increases and peaks near the SRT. A closer inspection of the low-field region [Fig. ~\ref{fig3}(d)] reveals a rapid decrease of MR in the vicinity of the SRT, without any sign reversal to positive MR. This pronounced low-field response is consistent with changes in the spin configuration and domain evolution close to the SRT.

Figure~\ref{fig3}(e) presents the field-dependent MR at $T=100~\mathrm{K}$ for different angles $\theta$ between $H$ and $I$, with the measurement geometry schematically illustrated in the inset of Fig. ~\ref{fig3}(f). In the low-field window ($-0.5~\mathrm{T} \le \mu_0H \le 0.5~\mathrm{T}$), MR is maximal for $\theta = 0^\circ$ ($\sim 0.7\%$) and decreases monotonically with increasing $\theta$ [Fig.~\ref{fig3}(f)]. For $\theta \ge 80^\circ$, MR becomes negative and reaches $\sim -0.3\%$ at $\theta = 90^\circ$ at $\pm 0.5~\mathrm{T}$. This pronounced angular dependence reflects a strong coupling between charge transport and the magnetic configuration near the spin-reorientation transition. As the magnetic field is rotated from the $ab$ plane toward the $c$ axis, drives the system toward a less-canted state, which reduces carrier scattering and results in a negative MR within $|\mu_0H| \lesssim 0.5~\mathrm{T}$. We ascribe this angular evolution to field-driven modifications of spin canting and magnetic domain configurations near $100~\mathrm{K}$.

\begin{figure*}
\centering
\includegraphics[width=1\textwidth]{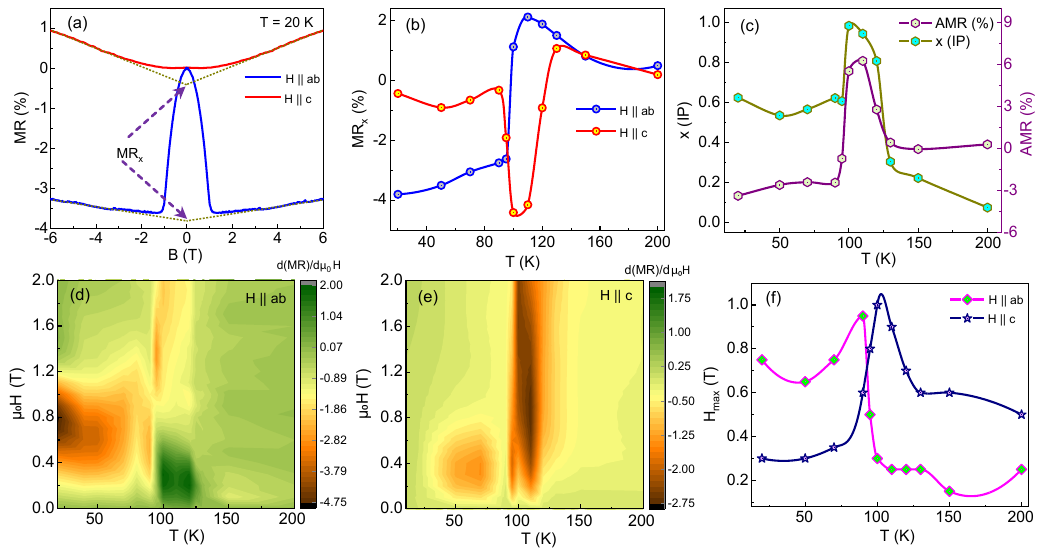}
\caption{\textbf{Anisotropic magnetoresistance (AMR) and inferred domain dynamics.} (a) The MR for the applied magnetic field along $ab$ the plane (blue) and $c$-axis (red) at 20 K. (b) Evolution of $MR_x$ with temperature for the two different magnetization directions. (c) The volume fraction of in-plain domains $x$ is shown on the left axis. The right axis shows the $AMR$ ratio as a function of temperature. (d-e) The contour plot of $\frac{d(MR)}{d(\mu_0H)}$ for H$\parallel ab$ and H$\parallel c$ as a function of $H$ and $T$. (f) Observed $T$ dependence of the field $H_{max}$ (where the field derivative $\frac{d(MR)}{d(\mu_0H)}$ shows a maximum).}
\vspace{-0.45cm}
\label{fig4}
\end{figure*}

\subsection{Anisotropic magnetoresistance (AMR)}

We now turn to the low-temperature MR data. At $T=20~\mathrm{K}$, the MR is positive for a magnetic field applied along the $c$ axis, whereas it is negative when the field is applied within the $ab$ plane, as shown in Fig.~\ref{fig4}(a). This pronounced AMR can arise from a field-direction dependence of the carrier mobility and/or from SOC effects in the itinerant electronic states \cite{Singh_2020,doi:10.1126/sciadv.aao3318}. To quantify the AMR, we define the resistivity of a fully magnetized state extrapolated to zero external field as $\rho_x$, where $x$ denotes the magnetization direction. The corresponding MR is
\begin{equation}
MR_x = \frac{\rho_x-\rho_0}{\rho_0},
\end{equation}
where $\rho_0$ is the zero-field resistivity, and $\rho_x$ is obtained by extrapolating the high-field resistivity to $\mu_0H \rightarrow 0$, as illustrated in Fig.~\ref{fig4}(b). The quantities $\rho_x$ and $MR_x$ capture the anisotropic resistivity governed by the magnetization orientation, magnetic-domain configuration, and spin-orbit coupling \cite{doi:10.1126/sciadv.aao3318}.

In the temperature interval $20$--$90~\mathrm{K}$, the magnitude of $|MR_c|$ is substantially smaller than that measured for the $H \parallel ab$ configuration, indicating that the $c$ axis is the magnetic easy axis in this regime. By contrast, above $\sim 90~\mathrm{K}$, $|MR_{ab}|$ becomes smaller than $|MR_c|$, consistent with preferential spin and domain alignment within the $ab$ plane, i.e., an easy-axis reorientation toward the basal plane at elevated temperatures. At all temperatures where $MR_x$ remains nonzero, the sample necessarily contains multiple magnetic-domain orientations, including a sizable fraction aligned along the preferred direction. To estimate the fraction of in-plane domains, we use the zero-field anisotropic resistivities $\rho_x$ within an effective-medium description \cite{PhysRevB.100.214420},
\begin{equation}
(1-x)\rho_{c} + x\rho_{ab} = \rho_{0},
\label{eq1}
\end{equation}
where $x$ is the volume fraction of domains whose magnetization lies in the $ab$ plane and $\rho_{0}$ is the measured zero-field resistivity. Using Eq.~\ref{eq1}, we determine $x(T)$ over the temperature range $20$--$200~\mathrm{K}$, as shown in Fig.~\ref{fig4}(c). The in-plane volume fraction exhibits its most pronounced change near $\sim 100~\mathrm{K}$, consistent with the spin-reorientation transition where spins and domains preferentially align in the basal plane. Upon cooling below this temperature, $x$ decreases toward $\sim 50\%$, indicating a progressive reorientation of spins and domains from the $ab$ plane toward the $c$ axis. Moreover, the zero-field resistivity $\rho(0)$ reflects the distribution of moments and domains and therefore lies between $\rho_{ab}$ and $\rho_c$. For the current applied in the $ab$ plane, we define the AMR ratio as $MR_{ab}-MR_c$. As shown in Fig.~\ref{fig4}(c), the AMR is maximized near $100~\mathrm{K}$, consistent with the strongest spin reorientation occurring in this temperature window.

A qualitative understanding of the temperature evolution of $MR_x$ can be obtained by considering the SRT and the accompanying redistribution of magnetic-domain volume fractions. Below $T_{\mathrm{SRT}}$, an applied field can drive domain reorientation between the basal plane and the $c$ axis, which modifies the effective resistivity and yields opposite MR responses for the two field geometries (negative and positive $MR_x$ for the out-of-plane and in-plane configurations, respectively). Importantly, the extracted domain volume fractions change most strongly in the vicinity of $T_{\mathrm{SRT}}$, indicating that the domains realign toward the preferred magnetization direction during the transition. The sign change of $MR_x$ observed near $\sim 100~\mathrm{K}$ in both configurations is therefore naturally associated with the completion of the SRT and the concomitant change of the magnetic easy axis. Even below $\sim 100~\mathrm{K}$, a finite fraction of the sample remains in domains aligned along the $c$ axis.

Over the full temperature range, the field dependence of MR differs markedly between $H \parallel ab$ and $H \parallel c$. The resulting two-stage evolution is more clearly resolved in the field derivative, $\frac{d(MR)}{d(\mu_0H)}$, evaluated at different temperatures. Figures~\ref{fig4}(d) and \ref{fig4}(e) show contour maps of $\frac{d(MR)}{d(\mu_0H)}$ as a function of $H$ and $T$ for the two field orientations. For $H \parallel ab$, $\frac{d(MR)}{d(\mu_0H)}$ is positive near $T_{\mathrm{SRT}}$ at low fields ($\mu_0H \sim 0.4~\mathrm{T}$). At low temperature and low field, $\frac{d(MR)}{d(\mu_0H)}$ becomes strongly negative, reflecting the rapid initial decrease of MR. Outside these regimes, the derivative is comparatively small and is further suppressed at higher fields, consistent with the observed two-stage field dependence. In contrast, the $H \parallel c$ geometry exhibits qualitatively different behavior in both field and temperature. Here, the most negative values of $\frac{d(MR)}{d(\mu_0H)}$ occur near $T_{\mathrm{SRT}}$, and a pronounced low-field response is also present around $50~\mathrm{K}$. This feature is more strongly suppressed upon increasing the field. The distinct evolution of $\frac{d(MR)}{d(\mu_0H)}$ for the two geometries points to multiple field- and temperature-dependent spin and domain configurations in F4GT. In the following section, we discuss the implications of these results.

The MR primarily reflects field-dependent scattering of itinerant carriers from magnetic domains and from spin fluctuations in F4GT. Here, we write the total MR as the sum of two contributions~\cite{Chakravorty2015,PhysRevB.85.094504},
\begin{equation}
 MR(H,T)=MR_{\mathrm{dom}}(H,T)+MR_{\mathrm{sf}}(H,T),
\end{equation}
where $MR_{\mathrm{dom}}$ originates from the field-driven evolution of magnetic domains (including domain-wall motion and domain reorientation), and $MR_{\mathrm{sf}}$ arises from the suppression of thermally excited spin fluctuations by an applied magnetic field. In the low-field regime, only modest fields are required to modify the domain configuration; hence $MR_{\mathrm{dom}}$ is expected to dominate. By contrast, at sufficiently high fields where the domains are largely aligned, the MR is governed mainly by the reduction of spin-disorder scattering and $MR_{\mathrm{sf}}$ becomes the leading contribution.

As a simple phenomenological description of the domain contribution, we adopt \cite{Chakravorty2015}
\begin{equation}
 MR_{\mathrm{dom}} \propto N_d\,\delta\,\xi(H),
\end{equation}
where $N_d$ is the number of domains that can be reoriented by the applied field, $\delta$ characterizes the resistivity change associated with domain rearrangement, and
\begin{equation}
 \xi(H)=\int_{0}^{H}\phi(\epsilon)\,d\epsilon .
\end{equation}
Here $\phi(\epsilon)$ is a distribution function of local fields $\epsilon$ that represent the effective anisotropy fields controlling domain reorientation, e.g., near grain boundaries or pinning centers. The distribution can be characterized by a most probable anisotropy field $H_A$. In this framework, the low-field derivative satisfies
\begin{equation}
 \frac{d(MR)}{d(\mu_0H)} \propto \phi(H),
\end{equation}
so that peaks or enhanced structure in $\frac{d(MR)}{d(\mu_0H)}$ directly reflect the underlying distribution of local anisotropy fields that govern domain motion.

Consistent with this picture, the MR shows its most pronounced low-field anomalies in the vicinity of the SRT, where the change in the in-plane domain fraction is maximal. In particular, near $T_{\mathrm{SRT}}$ we observe a small positive MR at low fields for $H \parallel ab$ [Fig.~\ref{fig3}(b)], accompanied by a pronounced peak in $\frac{d(MR)}{d(\mu_0H)}$, indicative of enhanced domain rearrangement controlled by the local anisotropy landscape. In this regime, the applied field can stabilize a domain configuration that increases the resistivity, yielding a positive low-field MR. Away from this narrow temperature window, the MR is predominantly negative. For $T<T_{\mathrm{SRT}}$, the buildup of magnetocrystalline anisotropy energy and the suppression of spin-disorder scattering favor a negative MR. For $T>T_{\mathrm{SRT}}$, the MR remains negative as both domain-related contributions and spin-fluctuation scattering diminish upon approaching $T_C$.

The derivative $\frac{d(MR)}{d(\mu_0H)}$ typically exhibits a broad extremum, from which we define a characteristic field $H_{\max}$. The temperature dependence of $H_{\max}$ is shown in Fig.~\ref{fig4}(f). Below $T_{\mathrm{SRT}}$, $H_{\max}$ for $H \parallel ab$ is slightly larger than for $H \parallel c$, consistent with an out-of-plane easy axis in this temperature range. Above $T_{\mathrm{SRT}}$, the trend reverses, indicating that the easy axis reorients into the $ab$ plane. At higher fields, MR approaches a limiting behavior consistent with a regime in which the applied field exceeds the characteristic field for domain alignment; consequently, $MR_{\mathrm{dom}}$ saturates and the remaining field dependence is governed primarily by the suppression of thermally excited spin fluctuations.

\begin{figure*}
\centering
\includegraphics[width=1.0\linewidth]{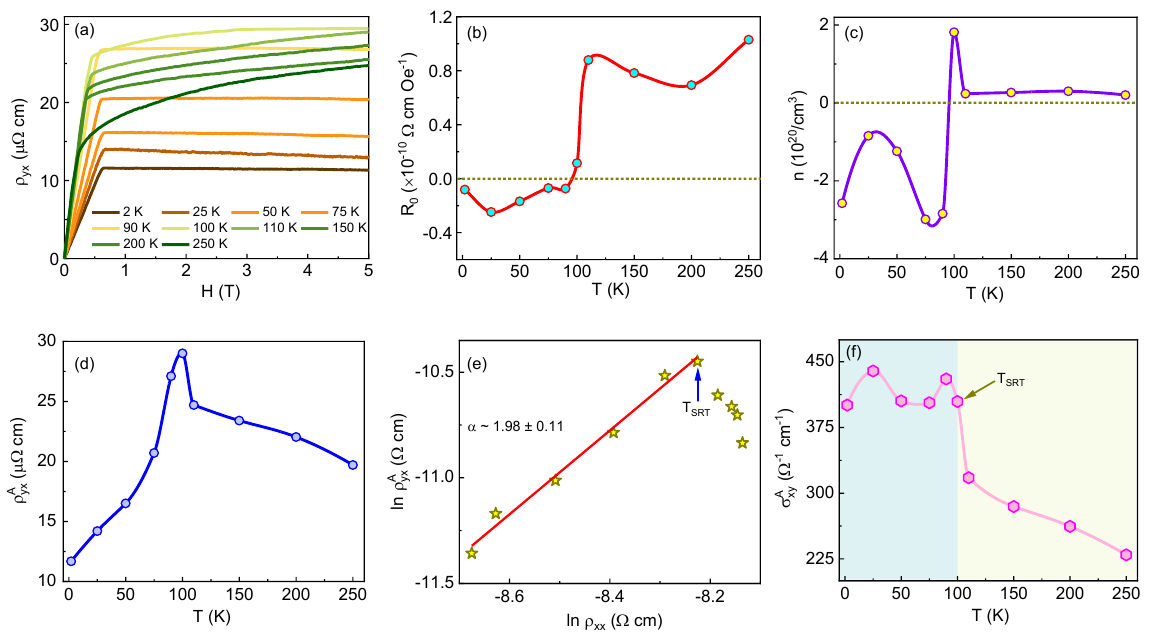}
\caption{\textbf{Hall resistivity and the anomalous Hall effect.} (a) Magnetic field-dependence of Hall resistivity ($\rho_{yx}$) at different temperatures. (b) Temperature dependence of the ordinary Hall coefficient (R$_0$). (c) Temperature-dependent carrier density $n$. (d) Anomalous Hall resistivity ($\rho^{A}_{yx}$) as a function of temperature. (e) Plot of ln $\rho^{A}_{yx}$ vs ln $\rho_{xx}$; the solid red line is the fit using the relation $\rho^{A}_{yx}$ $\propto$ $\rho^{\alpha}_{xx}$. (f) Temperature dependence of anomalous Hall conductivity ($\sigma^{A}_{xy}$).}
\label{fig5}
\end{figure*}

\subsection{Anomalous Hall effect}

Having established that the magnetotransport of F4GT is strongly affected by field-driven changes in spin and domain configurations through the MR response, we next measure the temperature-dependent Hall resistivity ($\rho_{yx}$) to identify how the corresponding evolution of carrier properties and possible Fermi-surface changes correlates with the MR anomalies across the SRT. Figure~\ref{fig5}(a) shows the Hall resistivity $\rho_{yx}$ as a function of magnetic field at representative temperatures. The low-field response exhibits a pronounced increase in $\rho_{yx}$ with a small applied field, and an anomalous contribution persists up to $\sim 0.6~\mathrm{T}$. At higher fields, $\rho_{yx}$ becomes only weakly field dependent and is well described by an approximately linear behavior. Notably, the overall character of the $\rho_{yx}(H)$ curves changes across the transition region in both the low- and high-field regimes.

In an FM, the Hall resistivity includes an ordinary contribution ($\rho^{0}_{yx}$) from the Lorentz force and an anomalous contribution ($\rho^{A}_{yx}$) that scales with the magnetization \cite{RevModPhys.82.1539},
\begin{equation}
\rho_{yx}(H)=\rho^{0}_{yx}(H)+\rho^{A}_{yx}(H)=R_0\,H+R_s\,\mu_{0}M,
\end{equation}
where $R_0$ and $R_s$ are the ordinary and anomalous Hall coefficients, respectively, and $\rho^{0}_{yx}$ and $\rho^{A}_{yx}$ denote the corresponding components of the Hall resistivity. We extract $R_0$ and $\rho^{A}_{yx}$ by fitting the high-field $\rho_{yx}(H)$ data to a linear form, where the slope yields $R_0$ and the intercept provides an estimate of $\rho^{A}_{yx}$ \cite{sudiptaCCGPhysRevB.107.125138,sudiptaPhysRevB.108.205108}. The resulting temperature dependence of $R_0$ is summarized in Fig.~\ref{fig5}(b), which reveals a clear sign change from positive to negative near $T_{\mathrm{SRT}} \approx 100~\mathrm{K}$. The sign reversal of $R_0$ therefore indicates a crossover of the dominant carriers from hole-like above $T_{\mathrm{SRT}}$ to electron-like below $T_{\mathrm{SRT}}$. Within a single-band approximation, the carrier density and dominant carrier type can be inferred from $R_0 = (ne)^{-1}$, where $e$ is the electron charge and $n$ is the carrier density. The extracted carrier density $n(T)$ shows a pronounced anomaly around $\sim 100~\mathrm{K}$ [Fig.~\ref{fig5}(c)], which reflects a reconstruction of the Fermi surface associated with the SRT. Figure~\ref{fig5}(d) displays the temperature dependence of the anomalous Hall resistivity, $\rho^{A}_{yx}$. A pronounced change is observed near $T_{\mathrm{SRT}}$, indicating a substantial modification of the anomalous Hall response across the SRT. Since $\rho^{A}_{yx}$ is sensitive to both the magnetic configuration (via $M$ and domain structure) and the underlying electronic structure (including Berry-curvature and scattering contributions), this anomaly suggests that the SRT is accompanied by a significant change in the effective Berry-curvature-driven transport and/or carrier scattering, consistent with a reconstruction of the low-energy electronic states in the vicinity of $T_{\mathrm{SRT}}$.

In order to understand how the anomalous Hall response evolves across the SRT, we analyze the scaling between $\rho^{A}_{yx}$ and $\rho_{xx}$ over the full temperature range using a double-logarithmic representation, as shown in Fig.~\ref{fig5}(e). Within the empirical scaling relation $\rho^{A}_{yx} \propto \rho_{xx}^{\alpha}$ \cite{RevModPhys.82.1539}, the exponent $\alpha$ provides a standard diagnostic of the dominant AHE mechanism: $\alpha \sim 1$ is characteristic of skew scattering, whereas $\alpha \sim 2$ indicates an intrinsic Berry-curvature contribution and/or the side-jump mechanism \cite{RevModPhys.82.1539}. A clear linear trend is obtained upto $T_{\mathrm{SRT}}$ in Fig.~\ref{fig5}(e), yielding $\alpha \simeq 1.98 \pm 0.11$ between 2 and 100~K, which evidences a Berry-phase-driven AHE in F4GT, consistent with previous reports \cite{Pal2024,PhysRevB.108.115122}. In contrast, the scaling relation breaks down above $T_{\mathrm{SRT}}$, where $\rho^{A}_{yx}$ can no longer be described by a single power law. This deviation suggests that the AHE is no longer governed by a nearly temperature-independent Berry-curvature contribution associated with a fixed Fermi-surface geometry, but instead reflects a substantial reorganization of the electronic topology and Berry curvature distribution. To further quantify this evolution, we convert $\rho^{A}_{yx}$ into the anomalous Hall conductivity using $\sigma^{A}_{xy} \approx \rho^{A}_{yx}/\rho_{xx}^{2}$ \cite{Chatterjee_2023}, and plot $\sigma^{A}_{xy}$ as a function of temperature in Fig.~\ref{fig5}(f). Notably, $\sigma^{A}_{xy}$ remains nearly temperature independent up to $T_{\mathrm{SRT}}$, as expected for an intrinsic AHE dominated by Berry curvature ``hot spots'' near the Fermi level \cite{sudiptaCCGPhysRevB.107.125138,Chatterjee_2023}. However, immediately above $T_{\mathrm{SRT}}$ we observe a pronounced drop in $\sigma^{A}_{xy}$, implying that the net Berry curvature contributing to the anomalous Hall response is strongly suppressed above $T_{\mathrm{SRT}}$. Such a suppression naturally follows from a Fermi-surface reconstruction associated with the reorientation of the magnetization axis, which can modify the spin--orbit--coupling-induced band splittings and the distribution of Berry curvature across the Brillouin zone. Above $T_{\mathrm{SRT}}$, the overall Hall response highlights that the SRT acts as an internal control knob that reshapes the Fermi-surface topology and the Berry-curvature landscape in F4GT.

\section{DISCUSSION}

Our magnetotransport data indicate that the SRT in F4GT is accompanied by pronounced changes in carrier scattering and transverse response, beyond a simple shift of magnetic anisotropy. The weak uniaxial anisotropy and small anisotropy field imply that modest thermal energies and applied fields can substantially modify the domain population and the degree of spin canting of the itinerant Fe moments. Correspondingly, $\rho_{xx}(T,H)$ becomes particularly sensitive to the field near $T_{\mathrm{SRT}}$, where we observe the most prominent low-field MR features, consistent with enhanced scattering from a field-evolving spin texture and reconfiguring domains. The extracted in-plane domain fraction $x(T)$ varies most rapidly around $T_{\mathrm{SRT}}$, supporting the view that the transition involves a redistribution of domain orientations rather than a purely coherent rotation of a single-domain magnetization. On the other hand, the Hall measurements provide a complementary perspective: the ordinary Hall coefficient $R_0(T)$ changes sign near $T_{\mathrm{SRT}}$, and the anomalous Hall resistivity $\rho^{A}_{yx}(T)$ exhibits a pronounced anomaly near SRT. Because $\rho^{A}_{yx}$ depends on both magnetic configuration (magnetization direction, canting, and domains) and the underlying electronic structure (through intrinsic Berry-curvature contributions \cite{PhysRevB.108.115122}), these features suggest that the SRT is accompanied by a substantial modification of the effective Hall response, potentially involving both magnetization-direction effects and changes in the low-energy electronic states. In this framework, it is plausible that the same reorientation of spins/domains that maximizes the MR anisotropy also contributes to the observed evolution of the anomalous Hall response.

The observed findings can also be discussed in the context of a recent theory proposing nearly flat bands in F4GT \cite{wang2023flat}. In that work, Fe$_{n}$GeTe$_2$ ($n=4,5$) is modeled as a stacked bipartite crystal lattice in which an imbalance of orbital degrees of freedom between sublattices can generate (nearly) flat bands; first-principles calculations further suggest that such bands may lie close to the $E_F$ and enhance the DOS, potentially favoring itinerant ferromagnetism in a Stoner-like picture. If F4GT indeed hosts near-$E_F$ flat-band-derived features, then its transport properties may be unusually sensitive to modest perturbations such as changes in exchange splitting, SOC-enabled hybridizations, and small shifts of chemical potential. In particular, because the magnetization direction can influence SOC-induced band mixing and avoided crossings, a spin reorientation could modify the Berry-curvature-driven intrinsic anomalous Hall response without requiring a large change in the magnitude of $M$. Likewise, a sign change in $R_0$ is compatible with a redistribution of multiband contributions to the ordinary Hall effect and could arise from temperature-dependent changes in the Fermi-surface topology of electron-like and hole-like bands; such effects could be amplified if the chemical potential lies near a sharp DOS feature. While our transport data alone do not uniquely determine the microscopic origin, the coincidence of strong low-field MR anomalies, rapid evolution of the inferred domain fraction, and pronounced changes in $R_0$ and $\rho^{A}_{yx}$ near $T_{\mathrm{SRT}}$ supports a scenario in which spin/domain reorganization and a magnetization-direction-sensitive electronic structure jointly shape the magnetotransport across the SRT.

Furthermore, a recent nitrogen-vacancy (NV) magnetometry study on ultrathin Fe$_4$GeTe$_2$ flake directly images the spin reorientation and reveals a pronounced enhancement of the magnetic anisotropy in the vicinity of $T_{\mathrm{SRT}}$ \cite{wang2024imaging}. This independent probe shows that the easy axis rotates in the same temperature range where we observe clear anomalies in the magnetotransport. Altogether, these results support a common origin in the strong, temperature-dependent anisotropy associated with the SRT.

\section{CONCLUSION}
In summary, we establish a well-defined temperature-driven spin-reorientation transition (SRT) in vdW FM, Fe$_4$GeTe$2$ near 100 K, marked by a crossover of the magnetic easy axis from the $c$ axis to the $ab$ plane with increasing temperature, based on comprehensive magnetic, thermodynamic, and transport measurements. Consistently, the Seebeck coefficient exhibits a pronounced anomaly near $T_{\mathrm{SRT}}$, providing independent evidence for a modification of the DOS in the vicinity of the Fermi level. The MR exhibits strong anisotropy between $H \parallel ab$ and $H \parallel c$ and a distinct two-stage field evolution, reflecting a low-field regime dominated by field-driven spin and domain reconfiguration and a higher-field regime in which the response is governed by the suppression of spin-disorder scattering. Analysis of the zero-field anisotropic resistivities within an effective-medium framework reveals a rapid change in the in-plane domain fraction near $T_{\mathrm{SRT}}$, implying a redistribution of domain orientations across the transition. Hall measurements further demonstrate a pronounced evolution of both ordinary and anomalous contributions: the ordinary Hall coefficient $R_0$ reverses sign near $T_{\mathrm{SRT}}$, while the anomalous Hall resistivity $\rho^{A}_{yx}$ exhibits a sharp jump in the same temperature window, suggesting a substantial modification of the effective transverse transport accompanying the spin reorientation. Furthermore, the scaling analysis of $\rho^{A}_{yx}$ and the temperature dependence of $\sigma^{A}_{xy}$ demonstrates that the intrinsic Berry-curvature-driven mechanism below $T_{\mathrm{SRT}}$ is strongly modified above the transition.

Taken together, our findings demonstrate that the SRT in F4GT acts as a control knob for electronic topology: the rotation of the magnetic easy axis not only modifies scattering mechanisms but also reorganizes the Berry-curvature distribution of the underlying bands. This establishes a direct link between spin anisotropy and topological transport in a quasi-2D vdW FM. The ability to manipulate such transitions through temperature or external fields suggests a route to realize reconfigurable topological states and spin-controlled transport devices in 2D vdW magnets.

\section*{ACKNOWLEDGMENTS}

This work was supported by the (i) Department of Science and Technology, Government of India (CRG/2023/001100), and (ii) CSIR-Human Resource Development Group (HRDG) (03/ 1511/23/EMR-II). S.C. thanks Dr. Saheli Samanta for her inputs. S.B. thanks CSIR, Govt. of India, for the Research Fellowship with Grant No. 09/080(1110)/2019-EMR-I.
\bibliographystyle{apsrev4-2}
\bibliography{F4GT_MR}{}
\end{document}